\documentclass[letter]{aa} % for the letters
\usepackage{epsfig}
\usepackage{rotating}
\usepackage{natbib}
\usepackage{graphicx}

\newcommand{\ros}{{\sl ROSAT}}
\newcommand{\euve}{{\sl EUVE}}
\newcommand{\iso}{{\sl ISO}}
\newcommand{\vlt}{{\sl VLT}}
\newcommand{\sao}{{\sl SAO}}
\newcommand{\fors}{{\sl FORS1}}
\newcommand{\forsn}{{\sl FOcal Reducer/low dispersion Spectrograph}}
\newcommand{\vltn}{{\sl Very Large Telescope}}
\newcommand{\chan}{{\sl Chandra}}

\newcommand{\gsc}{{ GSC2}}
\newcommand{\atca}{{\sl ATCA}}
\def \psr{PSR J0108$-$1431}
\def\pssr{J0108$-$1431}

\begin{document}
\title{A possible optical counterpart to the old nearby 
pulsar J0108$-$1431
\thanks{Based on observations collected at ESO, Paranal, under Programme 65.H-0400(A) }
}

\author{R. P. Mignani\inst{1}
\and
 G. G. Pavlov\inst{2}
\and
O. Kargaltsev\inst{2,3}
}

   \institute{Mullard Space Science Laboratory, University College London, Holmbury St. Mary, Dorking, Surrey, RH5 6NT, UK\\
              \email{rm2@mssl.ucl.ac.uk}
\and Department of Astronomy and Astrophysics, Pennsylvania
State University, PA 16802, USA, \email{pavlov@astro.psu.edu} \and 
Department of Astronomy,
University of Florida, FL 32611, USA, \email{oyk100@astro.ufl.edu} }

\titlerunning{Optical counterpart of \psr}

\authorrunning{Mignani et al.}
\offprints{R. P. Mignani; rm2@mssl.ucl.ac.uk}

\date{Received ...; accepted ...}

\abstract{The multi-wavelength study of old ($>$100 Myr) radio pulsars
holds  the key  to understanding  the long-term  evolution  of neutron
stars, including the  advanced stages of neutron star  cooling and the
evolution   of  the   magnetosphere.    Optical/UV  observations   are
particularly useful for such studies because they allow one to explore
both  thermal  and  non-thermal  emission processes.   In  particular,
studying the optical/UV emission constrains temperature of the bulk of
the  neutron  star   surface,  too  cold  to  be   measured  in  X-ray
observations.}{Aim of this work is to identify the optical counterpart
of the very old (166 Myr) radio pulsar \pssr.}{We have re-analyzed our
original \vltn\  (\vlt) observations (Mignani  et al.\ 2003),  where a
very  faint  object  was  tentatively  detected  close  to  the  radio
position, near the edge of a field galaxy.}{We found that the backward
extrapolation of  the \psr\ proper motion recently  measured by \chan\
(Pavlov et al.\  2008) nicely fits the position  of this object. Based
on  that,  we  propose  it  as  a viable  candidate  for  the  optical
counterpart to \psr.  The object  fluxes ($U =26.4 \pm0.3$; $B \approx
27.9$; $V  \ge 27.8$)  are consistent with  a thermal spectrum  with a
brightness temperature of $\sim 9\times 10^4$ K (for $R = 13$\,km at a
distance  of  130 pc),  emitted  from the  bulk  of  the neutron  star
surface.}   {New  optical observations  are  required  to confirm  the
optical identification of \psr\ and measure its spectrum.}


   \maketitle

\section{Introduction}

The radio pulsar \pssr\ was discovered by Tauris et al.\ (1994) during
the Parkes Southern Pulsar Survey (Manchester et al.\ 1996).  \psr\ is
certainly one of  the closest pulsars. Its dispersion  measure of 2.38
pc cm$^{-3}$  (D'Amico et al.\ 1998),  the smallest known  for a radio
pulsar, corresponds to a distance of 130 pc, according to the Galactic
electron density model by Taylor \& Cordes (1993) (180 pc in the model
by Cordes  \& Lazio 2002).   The pulsar's period,  $P = 0.808$  s, and
period  derivative,  $\dot{P}  =  7.44  \times  10^{-17}$  s  s$^{-1}$
(D'Amico  et al.\  1998),  correspond  to the  spin-down  age $\tau  =
P/2\dot{P}  = 166$  Myr, rotational  energy loss  rate $\dot{E}  = 5.8
\times  10^{30}$ ergs  s$^{-1}$, and  surface magnetic  field  $B= 2.5
\times 10^{11}$ G.  \psr\ is thus one of the oldest non-recycled radio
pulsars  known   to  date.   In   particular,  its  position   in  the
$P$-$\dot{P}$  diagram falls  quite close  to the  pulsar  death line.
Indeed, with a 400 MHz luminosity of $0.15 d^{2}_{130}$ mJy kpc$^{2}$,
where $d_{130}$ is  the distance in units of  130 pc, PSR J0108$-$1431
is the second faintest radio pulsar.

Soon after its discovery, there  were several attempts to detect \psr\
 at other wavelengths.  In X-rays,  the pulsar was not detected in the
 \ros\ All Sky Survey, while  pointed UV, IR, and optical observations
 with \euve\  (Korpela \& Bowyer  1998), \iso\ (Koch-Miramond  et al.\
 2002), and  the \sao\ RAS  6m telescope (Kurt  et al.\ 2000)  did not
 reveal candidate  counterparts.  Mignani et al.\  (2003) measured the
 most   accurate  radio-interferometric   position  with   \atca\  and
 performed deep optical observations  with the \vltn\ (\vlt), but they
 did  not   claim  a  candidate  counterpart.    Recently,  \psr\  was
 identified in a 30 ks  exposure with the \chan\ ACIS detector (Pavlov
 et  al.\ 2008),  based  on its  position  close to  the \atca\  radio
 coordinates  and on  the  high ($>300$)  X-ray-to-optical flux  ratio
 deduced with the  aid of the \vlt\ images of  Mignani et al.\ (2003).
 The  pulsar's spectrum  can be  fitted with  either a  power-law with
 photon  index  $\Gamma\approx2.2$ (for  the  hydrogen column  density
 $N_{\rm  H}  =  7.3  \times  10^{19}$ cm$^{-2}$  estimated  from  the
 pulsar's  dispersion measure)  or  a blackbody  with temperature  $kT
 \approx 0.28$  keV and  emitting area of  $\sim 50  d_{130}^2$ m$^2$.
 The estimated  0.3--8 keV  luminosity, $\sim 10^{28}  d_{130}^2$ ergs
 s$^{-1}$, translates into the X-ray emitting efficiency $\eta_X\equiv
 L_X/\dot{E} \sim  4 \times  10^{-3} d_{130}^2$, somewhat  higher than
 for most  of other  pulsars.  The comparison  between the  \chan\ and
 \atca\  coordinates   of  \psr\  yielded  its   first  proper  motion
 measurement   (199$\pm$65  mas   yr$^{-1}$),  corresponding   to  the
 transverse velocity of $129 \pm  42$ $d_{130}$ km s$^{-1}$ (Pavlov et
 al.\ 2008).

In this  paper we present a  re-analysis of the  \vlt\ observations of
 Mignani et  al.\ (2003).  The  analysis and results are  described in
 \S\,2, while the discussion is presented in \S\,3.

\section{Observations and data analysis}

Optical observations  of the  \psr\ field were  performed in  July and
August 2000 with  the \vlt, using the original  version of the \forsn\
(\fors).  At the time of  the observations the instrument was equipped
with a four-port 2048$\times$2084 CCD with a 0\farcs2 pixel size and a
field of view of $6\farcm8 \times 6\farcm8$ in its standard resolution
mode.  Observations  were performed in the  Bessel U, B  and V filters
(see  Mignani  et  al.\  2003   for  a  detailed  description  of  the
observations  and data reduction).   As discussed  by Mignani  et al.\
(2003), the identification of the \psr\ counterpart was complicated by
the fact  that the \atca\  radio position was  close to the edge  of a
relatively bright elliptical field galaxy.  A possible counterpart was
noticed in the  U and B bands only,  $\approx 0\farcs3$ northwest from
to the  radio position.  However, the  marginal detection significance
of this  object and its proximity  to the galaxy made  it difficult to
establish  whether  it  was   a  genuine  detection  or  a  background
enhancement, perhaps associated with  the galaxy.  For this reason, no
identification with the pulsar was claimed.

\subsection{Astrometry}

\begin{figure}
\centering
 \includegraphics[height=7.5cm,angle=0,clip]{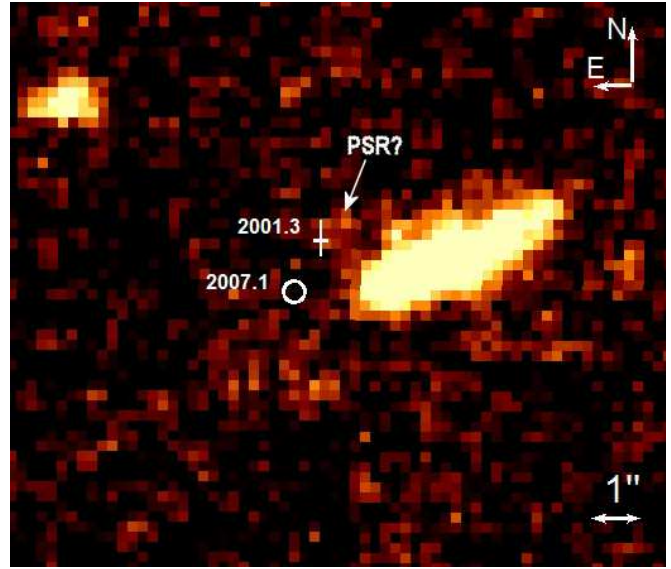}
  \caption{\vlt\ U band  image of the \psr\ field  obtained by Mignani
et al.\ (2003). The pulsar  candidate counterpart ($U = 26.4 \pm 0.3$)
is  marked by  an arrow.   The cross  shows  the radio-interferometric
position of Mignani et al.\ (2003) while the circle corresponds to the
most recent  pulsar position  obtained with \chan\  by Pavlov  et al.\
(2008).  The  position of  the  candidate  pulsar optical  counterpart
(epoch 2000.6) nicely fits the backward proper motion extrapolation.}
\label{fc}       % Give a unique label
\end{figure}

The recent pulsar  proper motion measurement by Pavlov  et al.\ (2008)
prompted  us  to re-investigate  the  possibility  that  \psr\ can  be
identified with the object marginally  detected in the \vlt\ data.  We
recomputed the astrometric calibration of our \vlt/\fors\ images using
as a  reference the  updated release of  the \gsc\ (ver.\  3.2), which
provided  a larger  number  of reference  stars  in the  field and  an
improved  coordinate accuracy, with  a mean  random radial  error (for
stellar  objects) $\sigma_{\rm GSC}=0\farcs3$  (Lasker et  al.\ 2008).
The   detector   coordinates  of   20   selected   \gsc\  stars   (all
non-saturated, with a well-defined PSF, and evenly distributed in the
field of view)  were measured by fitting a  Gaussian function to their
intensity  profiles  using  the  {\em Graphical  Astronomy  and  Image
Analysis}                         ({\em                         GAIA})
tool\footnote{http://star-www.dur.ac.uk/\~pdraper/gaia/gaia.html}.
The coordinate  transformation between the detector  and the celestial
reference  frame was then  computed using  the {\em  Starlink} package
{\tt ASTROM}\footnote{http://star-www.rl.ac.uk/Software/software.htm}.
The  rms of  our astrometric  fit was  0\farcs14 for  each of  the two
coordinates,   corresponding   to   the  radial   error   $\sigma_{\rm
fit}=0\farcs21$.  The new astrometric  solution is consistent with the
one obtained by Mignani  et al.\ (2003), but it is a  factor of 2 more
precise,  thanks  to the  larger  number of  \gsc\  stars  and to  the
improved astrometric  accuracy.  We estimated  the overall uncertainty
of our astrometric solution by adding in quadrature $\sigma_{\rm fit}$
and  the  uncertainty  $\sigma_{\rm  tr}  =  \sqrt{3/N_s}\,\sigma_{\rm
GSC}=0\farcs12$  with which  we can  register our  field on  the \gsc\
reference  frame  ($N_s=20$ is  the  number  of  reference stars,  and
$\sqrt{3}$ accounts  for the free  parameters in the  astrometric fit;
e.g., Lattanzi et  al.\ 1997).  The uncertainty on  the reference star
centroids is below 0\farcs01  and was neglected.  After accounting for
the $0\farcs15$ mean systematic uncertainty on the tie of the \gsc\ to
the International Celestial Reference Frame (Lasker et al.\ 2008), the
overall   uncertainty  of   our  absolute   astrometry   is  0\farcs28
($0\farcs19$ per coordinate).

\begin{figure*}
\centering
\includegraphics[bb=10 200 440 620,width=8.0cm,angle=0,clip=]{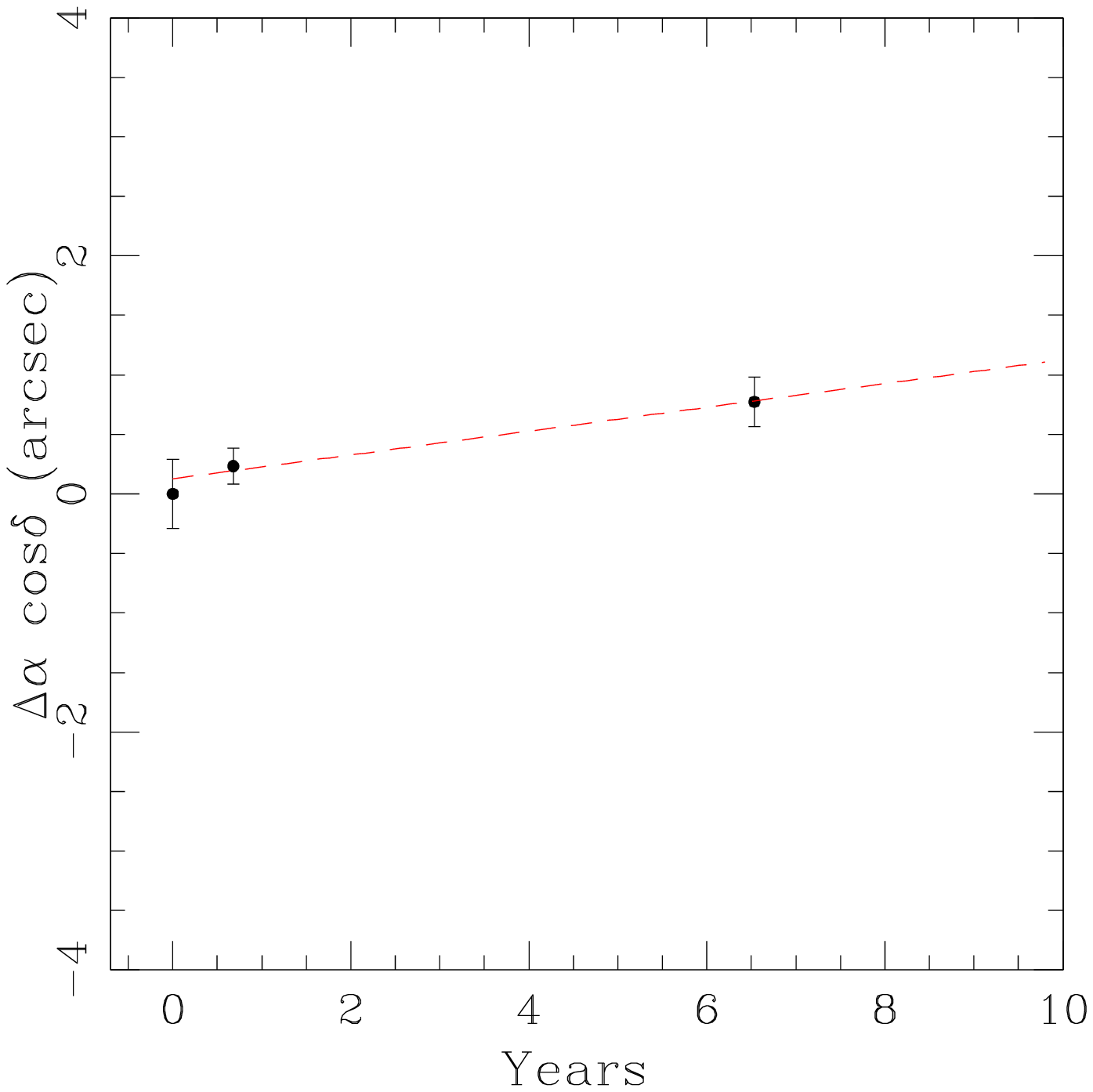}
\includegraphics[bb=10 200 440 620,width=8.0cm,angle=0,clip=]{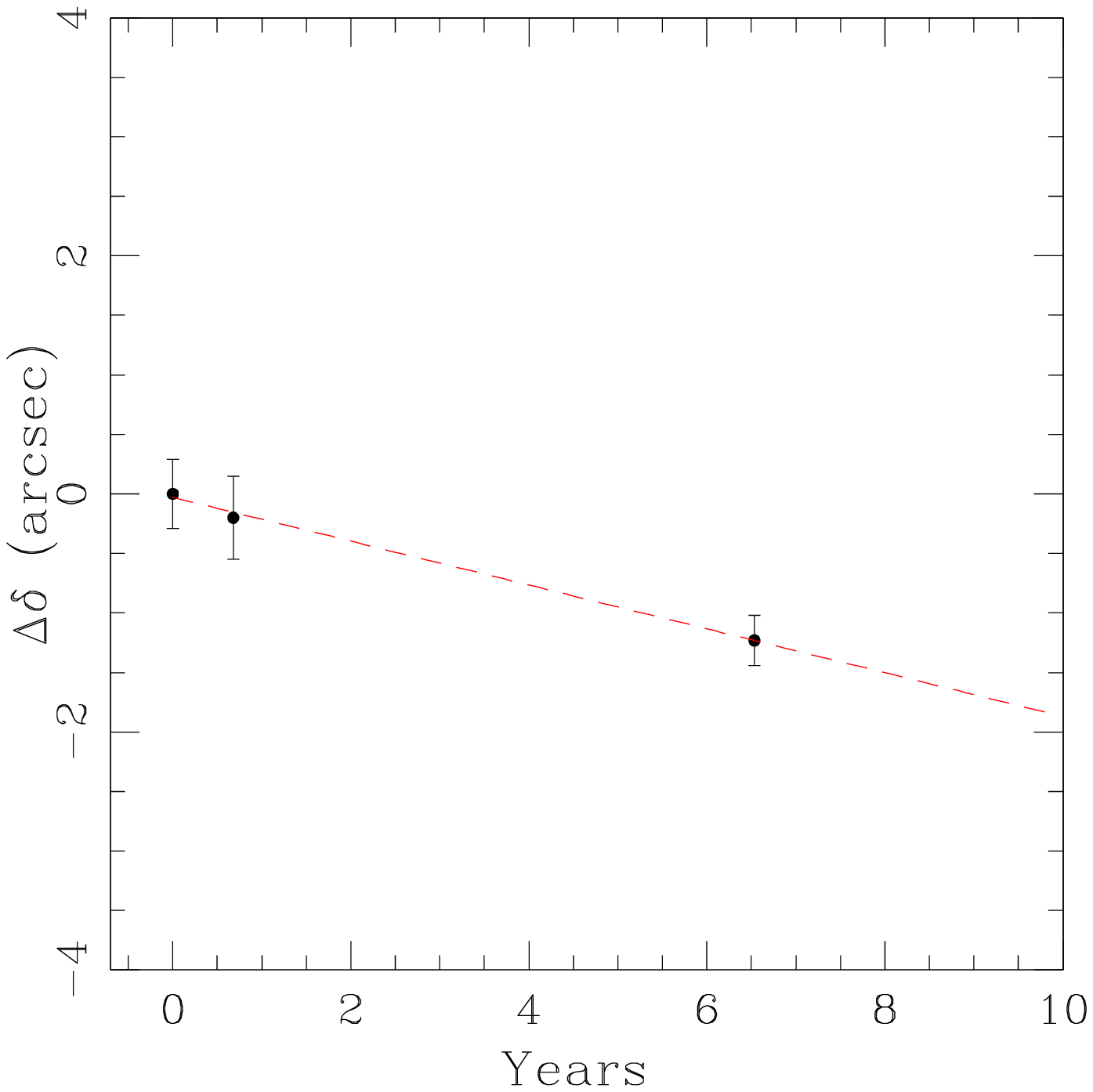}
\caption{Proper motion measurement in right ascension (left) 
and declination (right) computed from the optical (\vlt), radio (\atca), 
and X-rays (\chan) coordinates of \psr. The displacements are computed 
with respect to the first epoch \vlt\ position of the counterpart
candidate.} \label{pm}
\end{figure*}

The  \atca\  and \chan\  positions  of \psr\  are  shown  in Fig.\  1,
 overlayed on the \fors\ U band image after astrometric recalibration.
 As seen from Fig.\ 1, the backward proper motion extrapolation nicely
 fits the position  of the object tentatively detected  at the edge of
 the field galaxy. This is more quantitatively shown in Fig.\ 2, where
 we  compare  the relative  offsets  between  the  coordinates of  the
 optical source and those of  \psr\ measured by \atca\ and \chan.  The
 coordinates  of the  optical  source, computed  with our  astrometric
 solution, are $\alpha =01^{\rm h}  08^{\rm m} 08\fs301$ and $\delta =
 -14^\circ   31\arcmin  49\farcs15$,   (J2000),  with   the  $1\sigma$
 uncertainty of  $0\farcs27$ in each coordinate.   This value includes
 the  overall uncertainty of  our absolute  astrometry and  the object
 centroiding  error, added  in  quadrature.  Since  it is  practically
 impossible  to  fit  a  PSF  to  the object  profile,  we  assumed  a
 conservative  estimate of  $\pm  1$ pixel  ($\pm  0\farcs2$) for  the
 centroiding error.

  The yearly displacement computed using all the available coordinates
yields the proper motion of
$\mu_{\alpha} = 100 \pm 41$ mas yr$^{-1}$, $\mu_{\delta} = -184 \pm 49$
mas yr$^{-1}$
in right ascension and  declination, respectively. These proper motion
values are  in excellent  agreement with those  reported by  Pavlov et
al.\  (2008).
Thus,  based  on the
agreement between its position and the proper motion extrapolation, we
propose  the  object originally  identified  in  the  \vlt\ images  of
Mignani et al.\ (2003) as the candidate optical counterpart of \psr.

\subsection{Photometry}

We  carefully remeasured the  flux of  the candidate  counterpart. The
object is fairly well detected in the U band ($\approx 5 \sigma$), but
it is only marginally detected in the B band ($\approx 3 \sigma$). Due
to the difficulties in fitting a  model PSF to the object profile, we
measured  its flux  through aperture  photometry.  Sky  background was
sampled  in a  number of  regions selected  within a  radius  of $\sim
4\arcsec$ from the  object (far enough from the  field galaxy and from
other relatively  bright objects in the  vicinity), with sizes  2 to 4
times larger than the FWHM  of the image PSF.  
In order to increase as  much as possible the signal--to--noise ratio,
photometry  was  computed   using  customized  aperture  diameters  of
0\farcs6 and 0\farcs7, i.e.  about the  size of the image PSF in the B
and U-band, respectively. Due  to the uncertainty in the determination
of  the object centroiding  (see \S  2.1), measurements  were iterated
centering  the  aperture  at   different  positions  around  the  best
centroiding  estimate.   Results  were  found  consistent  within  the
measurement uncertainties. Aperture  correction was then applied using
as  a reference  the fitted  growing curves  of  well-suited reference
stars selected in the field.
Airmass correction was applied using
the Paranal extinction  coefficients measured with \fors.  Photometric
calibration was  applied using  the mean, extinction  corrected, night
zero points\footnote{http://www.eso.org/observing/dfo/quality/FORS1/qc/qc1.html}.

We  obtained $U=26.4  \pm  0.3$ and  $B\approx27.9$.   As the  higher
background  in the  B band  caused by  the close-by  galaxy  makes our
measurement very uncertain, we will use a fiducial error $\pm 0.5$ for
the B  magnitude.  The  object is  undetected in the  V band.  We have
remeasured  the flux  upper limit  at  the location  of the  candidate
counterpart and 
obtained $V> 27.8$  at a $3\sigma$  level, consistent
with  the  limit reported  by  Mignani  et  al.\ (2003).  The  derived
spectral fluxes  are shown  in Fig.\  3.  Since the  $N_{\rm H}  = 7.3
\times  10^{19}$  cm$^{-2}$, estimated  from  the pulsar's  dispersion
measure, implies a reddening  of $\approx 0.05$ magnitudes, i.e.  much
smaller  than our  photometric errors,  we neglected  the interstellar
extinction correction.

\section{Discussion}
  
With  only one fairly  accurate measurement,  our photometry  does not
allow  us   to  determine  the   spectral  slope  for   the  candidate
counterpart.   The  measured  optical  fluxes are  compatible  with  a
thermal (Rayleigh-Jeans) spectrum emitted from the bulk of the neutron
star surface with the (brightness) temperature $T=(7$--$10)\times 10^4
(d_{130}/R_{13})^2$  K, where  $R_{13}$ is  the apparent  neutron star
radius  in  units of  13  km. An  example  of  the blackbody  spectrum
consistent  with the optical  fluxes of  the candidate  counterpart is
shown in Fig.\ 3.

The U and B band fluxes and the V band upper limit cannot be described
by  a   flat  or  decreasing  power-law   spectrum,  $F_{\nu}  \propto
\nu^{-\alpha}$ with $\alpha\geq 0$, such as expected for magnetospheric 
emission.
It could be described by  an increasing power-law,
but power-laws  with $\alpha<0$ have not  been observed so  far in the
optical-UV  spectra  of  rotation-powered  neutron stars  (see,  e.g.,
Mignani et al.\ 2007).  Thus, it is unlikely that the optical emission
is  purely  magnetospheric. Of  course,  we  cannot  rule out  that  a
magnetospheric spectral  component is present  at a level below  the V
band limit (Fig.\ 3).

The optical  spectral fluxes are inconsistent  with the extrapolations
of the power-law  and blackbody X-ray fits (Fig.\  3).  In the thermal
interpretation, the  X-ray emission comes from small,  hot polar caps,
whose  contribution to  the optical  spectrum is  negligible,  but the
optical emission may  come from the bulk of  the neutron star surface,
which is  too cold  to be seen  in X-rays.   If the X-ray  spectrum is
predominantly magnetospheric, then  the optical through X-ray spectrum
of  the  magnetospheric  emission  cannot  be described  by  a  simple
power-law, but it could be described  by a model in which the spectral
slopes are different in  the X-rays and optical (e.g., $\alpha_X=1.2$,
$\alpha_O=0$; see Fig.\ 3 for an example).

If  further  observations  confirm  that  the  object  is  the  pulsar
counterpart  with  a  predominantly  thermal optical-UV  spectrum  (as
suggested  by Fig.\  3), this  would have  important  implications for
understanding  the thermal  evolution  of old  neutron  stars and  the
processes in their interiors.  Current neutron star cooling models are
rather uncertain for ages greater  than $\sim$1 Myr because the effect
of possible (re)heating processes  is unclear.  Such processes include
the Joule  heating, heating caused by readjustment  of the equilibrium
neutron  star structure  in the  course of  its evolution,  or heating
caused by the  effects of superfluidity in the  neutron star interiors
(see Fern\`{a}ndez \& Reisenegger 2005,  and Schaab et al.\ 1999 for a
review).  Evidence for 
reheating has been found in the much older
recycled pulsar J0437$-$4715, which showed a brightness temperature of
$\sim 1\times 10^5$ K (Kargaltsev et al.\ 2004).  The measurement of a
similar  temperature  in \psr\  would  mean  that 
some heating processes
operate  in slowly  rotating, non-recycled  pulsars. (We  should note,
however, that knowing the distance to the pulsar would be necessary to
measure  the surface  temperature from  the flux  measurements  in the
Rayleigh-Jeans tail, so that measuring  the parallax of \psr\ is badly
needed.)

If further  observations find a  non-thermal component in  the optical
spectrum,  this would  be  important for  understanding the  long-term
evolution  of the  rotation-powered  magnetospheric emission.   Recent
observations indicate  that even very old pulsars  are still efficient
magnetospheric  emitters,  with  an  even  higher  fraction  of  their
spin-down power emitted in the X-ray and optical bands than in younger
pulsars (Kargaltsev  et al.\  2006; Zharikov et  al.\ 2006;  Pavlov et
al.\ 2008).  In particular, in  the optical/UV this is demonstrated by
the cases of the pulsars  B1929+10 (Mignani et al.\ 2002) and B0950+08
(Zharikov  et al.\  2004),  which exhibit  flat  optical spectra  with
optical  efficiencies $\eta_O\equiv  L_O/\dot{E}\sim  1\times 10^{-6}$
and $\sim 4\times 10^{-6}$,  respectively (Zavlin \& Pavlov 2004).  If
at least  part of  the detected emission  from the  pulsar counterpart
candidate were of a magnetospheric origin, then the optical efficiency
(e.g., $\eta_O\sim  2\times 10^{-5}d_{130}^{2}$ for  the flat spectrum
shown by the dotted line in  Fig.\ 3) would be rather high compared to
the younger  pulsars B1929+10  and B0950+08.  On  the other  hand, the
(distance-independent)  ratio of  the X-ray  and  optical efficiencies
($\eta_O/\eta_X\sim 0.005$ for the  above example) is within the range
found for other pulsars detected in the optical (see Fig.\ 6 of Zavlin \& Pavlov 2004).

\begin{figure}
\centering
\includegraphics[bb= 60 60 560 700,angle=90,width=9cm,clip=]{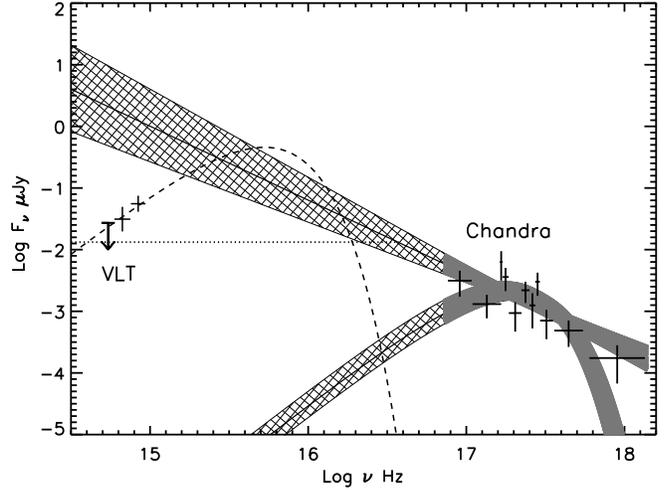}
\caption{Results of the optical photometry of the \psr\ candidate
counterpart (U and B band fluxes and V band upper limit) and 
the X-ray spectrum of \psr\ measured with \chan\ in the 0.3--8 keV band
(Pavlov et al.\ 2008).
The fits of the X-ray spectrum with the power-law and blackbody models
are shown, with their extrapolations to lower energies. The hatched
areas correspond to $1\sigma$ uncertainties of the X-ray fits.
The dashed curve shows a blackbody spectrum with the temperature of
$9\times 10^4$ K (for $R=13$ km, $d=130$ pc). 
The dotted line shows
an example of a flat power-law additional component in the optical-UV band, 
compatible with the measured V band upper limit.
}
\label{spectral}
\end{figure}

\section{Summary}

By re-analyzing  the \vlt\  images of Mignani  et al.\ (2003)  we have
identified {\em a posteriori}  a possible optical counterpart to \psr,
based on  its positional consistency  with the backward  pulsar proper
motion extrapolation.  If our identification is correct, this would be
the tenth (or  eleventh, if the detection of  PSR B1133+16 by Zharikov
et  al.\ 2008 is  confirmed) rotation-powered  pulsar detected  in the
optical/UV.  In any case, \psr\ would be the oldest non-recycled radio
pulsar  identified at  optical wavelengths.   As almost  8  years have
passed  since the  reported observations,  the  optical identification
could  be  easily  confirmed  by  proper  motion  measurement  of  the
candidate counterpart in deep optical/UV observations.

\begin{acknowledgements}
RPM  acknowledges  STFC  for  support  through  a  Rolling  Grant  and the
ESO/Chile Science  Visitor Programme for  supporting his visit  at the
Santiago offices, where this work was carried out.
\end{acknowledgements}


\begin{thebibliography}{99}

\bibitem[\protect\citeauthoryear{}{}]{} Cordes, J. M., \& Lazio, T. J. W., 2002, preprint (arXiv:astroph/0207156)

\bibitem[\protect\citeauthoryear{}{}]{} D'Amico, N., Stappers, B. W., Bailes, M., Martin, C. E., Bell, J.F., Lyne, A. G., \& Manchester, R. N., 1998, MNRAS, 297, 28

\bibitem[\protect\citeauthoryear{}{}]{} Kargaltsev, O., Pavlov, G.\ G., \& Romani, R.\ W., 2004, ApJ, 602, 327

\bibitem[\protect\citeauthoryear{}{}]{} Kargaltsev, O., Pavlov, G.\ G., \& Garmire, G.\ P., 2006, ApJ, 636, 436

\bibitem[\protect\citeauthoryear{}{}]{} Koch-Miramond, L., Haas, M., Pantin, E., Podsiadlowski, P., Naylor, T., \& Sauvage, M., 2002, A\&A, 387, 233

\bibitem[\protect\citeauthoryear{}{}]{} Korpela, E. J., \& Bowyer, S., 1998, AJ, 115, 2551

\bibitem[\protect\citeauthoryear{}{}]{} Kurt, V. G., Komarova, V. N., Fatkhullin, T. A., Sokolov, V. V., Koptsevich, A. B., \& Shibanov, Yu. A., 2000, Bull. Spec. Astrophys. Obs., 49, 5

\bibitem[\protect\citeauthoryear{Lasker et al.}{2008}]{las08} Lasker, B.M., et al., 2008, submitted to AJ

\bibitem[]{} Lattanzi, M.\ G., Capetti, A., \& Macchetto, F.\ D. 1997, A\&A, 318, 997

\bibitem[\protect\citeauthoryear{}{}]{} Mignani, R. P., De Luca, A., Caraveo, P. A., Becker, W., 2002, ApJ, 580, 147

\bibitem[\protect\citeauthoryear{}{}]{} Mignani, R. P., Manchester, R. N., \& Pavlov, G. G., 2003, ApJ, 582, 97

\bibitem[\protect\citeauthoryear{Mignani et~al.}{2007}]{mig07}Mignani, R.P., Zharikov,  S., Caraveo,  P.A., 2007, A\&A, 473, 891

\bibitem[\protect\citeauthoryear{}{}]{} Pavlov, G.\ G., Kargaltsev, O., Wong, J.\ A., \& Garmire, G.\ P., 2008, ApJ, submitted (arXiv:0803.0761)

\bibitem[]{} Fern\`{a}ndez, R., \& Reisenegger, A. 2005, ApJ, 625, 291

\bibitem[\protect\citeauthoryear{}{}]{} Schaab, Ch., Sedrakian, A., Weber, F., \& Weigel, M.\ K., 1999, A\&A, 346, 465

\bibitem[\protect\citeauthoryear{}{}]{} Tauris, T. M., et al., 1994, ApJ, 428, L53

\bibitem[\protect\citeauthoryear{}{}]{} Taylor, J. H., \& Cordes, J. M., 1993, ApJ, 411, 674

\bibitem[\protect\citeauthoryear{}{}]{} Zavlin, V.\ E., \& Pavlov, G.\ G., 2004, ApJ, 616, 452

\bibitem[\protect\citeauthoryear{}{}]{} Zharikov, S. V., Shibanov, Yu. A., Mennickent, R. E., Komarova, V. N., \& Tovmassian, G. H., 2004, A\&A, 417, 1017

\bibitem[\protect\citeauthoryear{}{}]{} Zharikov, S. V., Shibanov, Yu. A., \& Komarova, V. N., 2006, AdSpR, 37, 1979

\bibitem[\protect\citeauthoryear{}{}]{} Zharikov, S. V., Shibanov, Yu. A., Mennickent, R. E. \& Komarova, V. N., 2008, A\&A, 479, 793

\end{thebibliography}
\end{document}